\documentclass[aps,prb,reprint]{revtex4-1}

\usepackage{amsmath}
\usepackage{graphicx}
\usepackage{dcolumn}
\usepackage{verbatim}

\newcolumntype{d}{D{.}{.}{-1}}

\begin{document}

\title{Variationally fitting the total electron-electron interaction}
\author{Brett I. Dunlap}
\email{dunlap@nrl.navy.mil}
\author{Mark C Palenik}
\thanks{NRC Research Associate}
\address{Code 6189, Chemistry Division, Naval Research
	Laboratory, Washington, DC 20375, United States}

\begin{abstract}

Density fitting is used throughout quantum chemistry to simplify the electron-electron interaction energy (EE). A fundamental property of quantum chemistry, and DFT in particular, is that a variational principle connects the EE to a potential.  Density fitting generally does not preserve this connection. Herein, we describe the construction of a robust EE that is variationally connected to fitted potentials in all electronic structure methods. For DFT, this results in new fitting equations which are satisfied at an energy saddle point in multidimensional fitting space.

\end{abstract}

\maketitle

\begin{figure*}[t]
	\includegraphics[width=2\columnwidth]{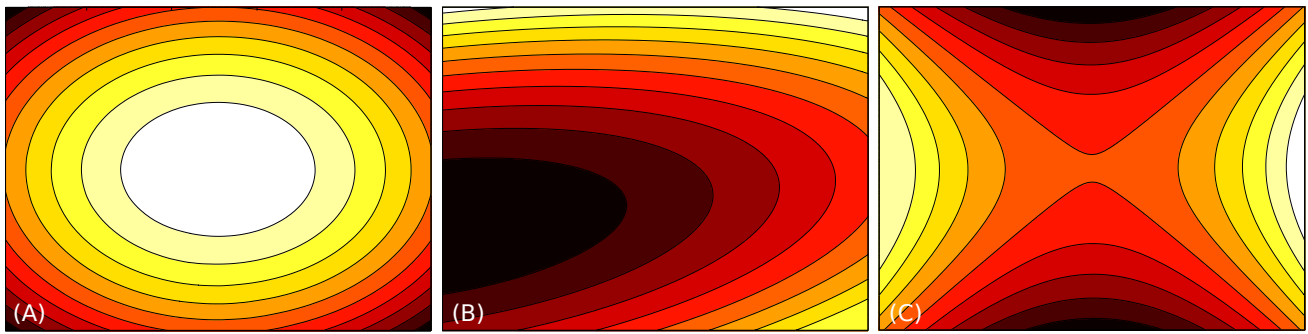}
	\caption{Robust Coulomb plus XC energy of Zn with VWN versus fitting coefficients.  Light high energy and dark is low. The plots are centered on the original SCF fit.  (a) and (b) are centered on the Coulomb fit and the axes are Coulomb matrix eigenvectors of the two largest (a) and smallest (b) eigenvalues.  The fitted density is used to compute XC.  (c) Centered on the variational EE fit.  The axes are eigenvectors of the variational EE [Eq.~(\ref{Eq2})] Hessian.}
	\label{Figure1}
\end{figure*}

Variational principles are of fundamental importance in physics.  The evolution of a physical system can be determined by applying the calculus of variations to a scalar quantity, known as the action.  In density functional theory (DFT), this is manifested through the Hohenberg-Kohn theorems \cite{HKTheorems}, which prove that the ground-state electron-density minimizes the total energy.  In Kohn-Sham (KS) DFT, the electron-electron interaction energy (EE), which includes Coulomb (Hartree) and exchange-correlation terms, is also variationally linked to the KS potential \cite{Kohn1965}.

Computation of the all-electron Gaussian-orbital-based energy in KS-DFT is only simplified relative to the equivalent Hartree-Fock (HF) calculations if the KS potential is fitted \cite{SambeFelton1975}, yielding methods that scale as the number of electrons to the third, rather than fourth, power.  The widely used variational Coulomb fitting method makes the Coulomb potential a variational function of a fitted electron-density \cite{Dunlap1979,Dunlap1979-2}.   This both simplifies computations and makes the Coulomb energy stationary with respect to the fitting coefficients \cite{Dunlap2010Mol}.

Fitting the Coulomb potential still leaves the question of how best to simplify computation of the remainder of the electron-electron interaction.  Early on, for two-dimensional periodic systems, the Coulomb-fitted electron-density was used to generate a second, non-variational, analytic fit to the exchange-correlation (XC) potential \cite{Boettger1985,Boettger1989}.  More recently, in molecular systems, the Coulomb-fitted density has been used directly to numerically calculate the XC potential \cite{Birkenheuer2005,Flores-Moreno2008}.  However, as can be seen by FIG.~\ref{Figure1}\,(b), the fit is not located at a stationary point of the EE.  No previously existing fitting method preserves the variational relationship between the EE and the KS potential.  

Our method allows us to variationally connect the EE of \textit{any} electronic-structure method to a single-center fitted KS-like potential or set of potentials.  For HF, it reproduces existing fitting equations \cite{Whitten1973,Vahtras1993} by construction, but now expressed as the calculus of variations applied to a variationally stable energy.  For DFT, simultaneously fitting the total Coulomb and XC energy results in new fitting equations, which we explore with calculations on transition metal atoms.  The new fitting metric, which includes Coulomb and XC terms, also naturally arises in and symmetrizes the equations for perturbation theory with density fitting \cite{Palenik2015}.

The variational link between the EE and the potential is the result of defining a new, robust energy, which contains no first-order errors due to fitting \cite{Dunlap2000}.  The robust EE includes {\it both} exact and fitted densities and can be made fully stationary with respect to both sets of variables.  Doing so generates a coupled set of orbital self consistent field (SCF) and variational density fitting equations.

The KS potential in KS-DFT is defined as the variation of the EE with respect to the density, where the density is given by a sum over orbitals squared.  This single potential can be generalized to encompass electronic structure methods in which the EE depends on multiple density-like quantities by defining a series of potentials that act on each quantity.

For example, in HF, the EE can be written in terms of local charge distributions given by the product of orbital pairs, $\rho_{ij}(\mathbf{r})=\phi_i^*(\mathbf{r})\phi_j(\mathbf{r})$.  We shall refer to each independent local function of orbital-pairs that appears in a given electronic structure method as a generalized density, labeled by a single index, $\rho_j(\mathbf{r})$.  For each $\rho_j(\mathbf{r})$, we can define a local potential $V_{j}(\mathbf{r})$
\begin{equation}
V_{j}(\mathbf{r})=\frac {\delta E_{ee}[\rho]} {\delta\rho_{j}(\mathbf{r})} 
\label{Eq1}
\end{equation}
where $E_{ee}[\rho]$ is the EE, written as a function of all of the generalized densities.

The potential $V_{j}(\mathbf{r})$ acts on $\rho_j(\mathbf{r})$.  However, the integral $\int V_{j}(\mathbf{r})\rho_j(\mathbf{r})d\mathbf{r}$ is not the corresponding contribution to the EE.  Rather, because the EE is nonlinear in the $\rho_j(\mathbf{r})$'s, a double-counting (DC) correction is required.

This DC correction plays an important role in defining the robust energy.  A fitted generalized density, $\bar{\rho}_j(\mathbf{r})$, will differ from the exact density if the fitting basis set is incomplete.  This fitting error means that $E_{ee}[\bar{\rho}]\neq E_{ee}[\rho]$ and therefore, there are may different ways we can write the EE using combinations of the the fitted and exact densities which are not equivalent.  If we make the requirements that the electron-electron potentials are purely in terms of the fitted density and that these potentials are produced by variation of $E_{ee}$ with respect to the exact densities, there is only one possible choice for $E_{ee}$
\begin{equation}
E_{ee}[\rho,\bar{\rho}]=E_{DC}[\bar\rho]+\sum_{j}\int V_{j}[\bar{\rho}(\mathbf{r})]\rho_{j}(\mathbf{r})d\mathbf{r}
\label{Eq2}
\end{equation}
where $E_{DC}$ is the double counting correction and is written entirely in terms of the fitted density.  This definition of the EE is robust, meaning that it contains a first order correction for the difference between each $\rho_j(\mathbf{r})$ and $\bar\rho_j(\mathbf{r})$.  It is easy to show this by explicitly writing the fitted EE and then adding the first term in the Taylor series that connects it to the exact EE, yielding
\begin{equation}
\begin{split}
E_{ee}[\bar{\rho}]+\sum_j\int\frac{\delta E_{ee}[\bar{\rho}]}{\delta\bar{\rho}_j(\mathbf{r})}(\rho_j(\mathbf{r})-\bar{\rho}_j(\mathbf{r}))d\mathbf{r}\\
=E_{ee}[\bar{\rho}] + \sum_j\int V_j[\bar{\rho}(\mathbf{r})](\rho_j(\mathbf{r})-\bar\rho_j(\mathbf{r}))d\mathbf{r}\\
=E_{DC}[\bar{\rho}]+\sum_j\int V_j[\bar{\rho}(\mathbf{r})]\rho_j(\mathbf{r}) d\mathbf{r}
\end{split}
\end{equation}
The variation of this energy with respect to $\rho_j(\mathbf{r})$ produces $V_j[\bar{\rho}(\mathbf{r})]$, which we will from now on refer to as $\bar V_j(\mathbf{r})$.

Setting the variation of the robust energy with respect to each $\bar\rho_j(\mathbf{r})$ to zero defines a coupled set of fitting equations
\begin{equation}
\begin{split}
\delta E_{ee}&=\mkern-5mu	\int\left[\frac{\delta
	E_{DC}[\bar\rho]}{\delta\bar\rho_{k}(\mathbf{r})}
+\sum_{j}	\int \frac{\delta
	\bar V_{j}(\mathbf{r'})} {\delta\bar\rho_{k}(\mathbf{r})}\rho_{j}(\mathbf{r'})d\mathbf{r'}\right]\mkern-5mu\delta\bar\rho_k(\mathbf{r})d\mathbf{r}\\
&=	\sum_{j}
\int \left[\rho_{j}(\mathbf{r'})-\bar\rho_{j}(\mathbf{r'})\right]\frac{\delta
	\bar V_{j}(\mathbf{r'})} {\delta\bar\rho_{k}(\mathbf{r})} \delta\bar{\rho}_k(\mathbf{r})d\mathbf{r}d\mathbf{r'}=0
\end{split}
\label{Eq3}
\end{equation}
where $\delta\bar{\rho}_k(\mathbf{r})$ represents any variation of $\bar{\rho}_k(\mathbf{r})$ allowed by our fitting procedure.  When $\bar{\rho}_k(\mathbf{r})$ is expanded in a fitting basis, the $\delta\bar{\rho}_k(\mathbf{r})$ will be basis functions and Eq.~(\ref{Eq3}) can be solved to find the fitting coefficients.  Equation~(\ref{Eq3}) is then the derivative of the robust EE with respect to the fitting coefficients.  Differentiating a second time, we get the EE Hessian, and its eigenvectors were used to create Fig.~\ref{Figure1}\,(c).

In general, Eq.~(\ref{Eq3}) can be used to determine any variational parameter, which includes the fitting coefficients, and in the case of basis set optimization, the basis-function exponents, in the same manner as has been described for the Coulomb-fitting equations \cite{Dunlap1979}.  In fact, if instead of using $E_{ee}$, we substitute in the Coulomb energy, $\bar{V}_j(\mathbf{r})$ becomes $\bar{\rho}_j(\mathbf{r})/|\mathbf{r}-\mathbf{r'}|$ and Eq.~(\ref{Eq3}) becomes
\begin{equation}
\int \frac{\rho(\mathbf{r'})-\bar\rho(\mathbf{r'})}{|\mathbf{r}-\mathbf{r'}|}
\delta\bar\rho_k(\mathbf{r})d\mathbf{r}d\mathbf{r'}=0.
\label{Eq5}
\end{equation}
Because the equations for each $\bar{V}_j(\mathbf{r})$ are identical, only the total density needs to be fitted, and Eq.~(\ref{Eq5}) is the Coulomb fitting equation.  This equation is linear in the fitted density $\bar{\rho}$, and therefore, fitting coefficients.  Unlike Eq.~(\ref{Eq5}), however, Eq.~(\ref{Eq3}) is in general, nonlinear because the XC potential is a nonlinear function of $\bar{\rho}$.

As we did with the total EE, we can find the Hessian of the robust Coulomb energy by differentiating the left hand side of Eq.~(\ref{Eq5}) with respect to fitting coefficients.  Because of the minus sign in front of $\bar{\rho}$, we are left with the negative of the Coulomb interaction between pairs of basis functions, or in other words, the negative of the Coulomb matrix, which itself is positive definite.  The contribution of the Coulomb energy to the eigenvalues of the robust EE Hessian is, therefore, negative.

A nice property shared by variational Coulomb fitting and our new variational fitting method is that they both make the total energy stationary, and not simply the portions related to electron-electron interactions.  Another way to arrive at Eq.~(\ref{Eq3}) (or Eq.~(\ref{Eq5}) when $\bar{\rho}$ is only used in the Coulomb energy) is by using perturbation theory to compute the first order change in energy when $\bar{\rho}$ becomes $\rho$ and setting it equal to zero.  This first order energy is the derivative of the total kinetic, electron-nuclear, and electron-electron energy with respect to the fitted density.  Variational fitting therefore truly makes the total energy stationary.  The same cannot be said when the Coulomb fitted density is inserted directly into XC.

The general DFT XC energy is a functional of the spin densities and their gradients, Laplacians, {\it etc}., up Perdew's ladder of functionals \cite{Perdew2005}.  For the second rung, the generalized gradient approximation (GGA), in addition to spin densities, the EE depends on $\sigma_1=|\nabla\rho_\uparrow(\mathbf{r})|^2$, $\sigma_2=|\nabla\rho_\downarrow(\mathbf{r})|^2$, and $\sigma_3=\nabla\rho_\uparrow(\mathbf{r})\cdot\nabla\rho_\downarrow(\mathbf{r})$, where the arrow indicates the direction of the spin.  There is one independent $V_j$ for each of these variables.  These potentials can be combined into two potential operators, one for each spin, that act on the corresponding spin density.  For spin up, the XC potential operator is
\begin{equation}
V_{XC\uparrow}=V_{\rho_\uparrow}(\mathbf{r})+\left[2V_{\sigma_1}(\mathbf{r})\nabla\rho_\uparrow(\mathbf{r})+V_{\sigma_2
}(\mathbf{r})\nabla\rho_\downarrow(\mathbf{r})\right]\cdot\nabla
\label{Eq4}
\end{equation}
with a similar expression for spin down.

All practical Gaussian fitting basis sets are incomplete, introducing an error that is significant for transition metal atoms \cite{Dunlap1977}.  Therefore, we chose Zn($^\mathrm{1}$S) (Table~\ref{Table1}) and paramagnetic Mn($^\mathrm{6}$S) with five unpaired spins (Table~\ref{Table2}) to test our new fitting method.  Moreover, these atoms are spherically symmetric and angular integration grids were not needed.

For comparison, we also performed calculations using the model-potential method of ParaGauss \cite{Birkenheuer2005} and the auxiliary density functional theory (ADFT) of deMon2K \cite{Flores-Moreno2008}.  In the model-potential of ParaGauss, the normalized, Coulomb-fitted density is used to compute the XC energy.   A Lagrange multiplier enforces normalization of the  fit.  The unconstrained version of this approach is ADFT.

\begin{table*}[t]
	\caption{Robust, relative atomic Zn energies.  For each functional, VWM, PBE,
		and BLYP, the Coulomb, CE, electron-electron, EE, and total energies, in
		Hartree, are given relative to SCF energies computed with unconstrained, variationally fitted
		CE and the XC energy of the true density.}
	\centering
	\begin{ruledtabular}
		\begin{tabular}{l c| d d d| d d d| d d d}
			&&\multicolumn{3}{c|}{VWN}&\multicolumn{3}{c|}{PBE}&\multicolumn{3}{c}{BLYP}\\
			\cline{3-11}
			Fitting Method&Constraint&\multicolumn{1}{c}{CE}& \multicolumn{1}{c}{EE}&\multicolumn{1}{c|}{Total}&\multicolumn{1}{c}{CE}&\multicolumn{1}{c}{EE}&\multicolumn{1}{c|}{Total}&\multicolumn{1}{c}{CE}&\multicolumn{1}{c}{EE}&\multicolumn{1}{c}{Total}\\  
			\hline
			Numeric& Yes&
			-0.002	&	-0.002	&	0.000	&	-0.002	&	-0.002	&	0.000	&	-0.002	&	-0.002	&	0.000\\
			\hline
			Variational&Yes&
			0.053	&	0.063	&	0.003	&	0.079	&	0.086	&	0.008	&	-0.222	&	0.248	&	-0.016\\
			Densities&No&
			0.053	&	0.063	&	0.003	&	0.070	&	0.076	&	0.008	&	-0.197	&	0.142	&	-0.006\\
			\hline
			Variational& Yes&
			-0.125	&	-0.112	&	0.004	&	-0.167	&	-0.150	&	0.008	&	-0.102	&	-0.091	&	0.005\\
			Coulomb&No&
			-0.056	&	-0.046	&	0.004	&	-0.108	&	-0.094	&	0.008	&	-0.047	&	-0.039	&	0.005\\
		\end{tabular}
	\end{ruledtabular}
	\label{Table1}
\end{table*}

\begin{table*}[t]
	\caption{Robust, relative atomic Mn energies.  For each functional, VWM, PBE,
		and BLYP, the Coulomb, CE , electron-electron, EE, and total energies, in
		Hartree, are given relative to SCF energies computed with unconstrained, variationally fitted
		CE and the XC energy of the true density.}
	\centering
	\begin{ruledtabular}
		\begin{tabular}{l c| d d d| d d d| d d d}
			&&\multicolumn{3}{c|}{VWN}&\multicolumn{3}{c|}{PBE}&\multicolumn{3}{c}{BLYP}\\
			\cline{3-11}
			Fitting Method&Constraint&\multicolumn{1}{c}{CE}& \multicolumn{1}{c}{EE}&\multicolumn{1}{c|}{Total}&\multicolumn{1}{c}{CE}&\multicolumn{1}{c}{EE}&\multicolumn{1}{c|}{Total}&\multicolumn{1}{c}{CE}&\multicolumn{1}{c}{EE}&\multicolumn{1}{c}{Total}\\ 
			\hline
			Numeric& Yes&
			0.000	&	0.000	&	0.000	&	0.000	&	0.000	&	0.000	&	0.000	& 	0.000	&	0.000\\
			\hline
			Variational&Yes&
			0.015	&	0.016	&	0.002	&	0.027	&	0.030	&	0.004	&	0.032	&	0.036	&	0.002\\
			Densities&No&
			0.014	&	0.016	&	0.002	&	0.027	&	0.029	&	0.004	&	0.012	&	0.018	&	0.002\\
			\hline	
			Variational& Yes&
			-0.011	&	-0.008	&	0.002	&	-0.037	&	-0.030	&	0.004	&	-0.029	&	-0.024	&	0.003\\
			Coulomb&No&
			0.002	&	0.004	&	0.002	&	-0.022	&	-0.016	&	0.004	&	-0.019	&	-0.014	&	0.003\\
			\hline
			Variational& Yes&
			-0.015	&	-0.003	&	-0.008	&	0.044	&	0.046	&	0.004	&	0.018	&	0.020	&	0.003\\
			Polarization&No&
			0.000	&	0.011	&	-0.007	&	0.051	&	0.052	&	0.004	&	0.024	&	0.025	&	0.003
		\end{tabular}
	\end{ruledtabular}
	\label{Table2}
\end{table*}

\begin{table*}[t]
	\caption{The first-order XC energy error in Hartree due to fitting for atomic Mn
		and Zn using the three functionals.}
	\centering
	\begin{ruledtabular}
		\begin{tabular}{l c| d d d| d d d}
			&&\multicolumn{3}{c|}{Mn}&\multicolumn{3}{c}{Zn}\\
			\cline{3-8}
			Fitting
			Method&Constraint&\multicolumn{1}{c}{VWN}& \multicolumn{1}{c}{PBE}&\multicolumn{1}{c|}{BLYP}&\multicolumn{1}{c}{VWN}& \multicolumn{1}{c}{PBE}&\multicolumn{1}{c}{BLYP}\\ 
			\hline
			Variational&Yes&
			0.006	&	-0.004	&	-0.031	&	0.007	&	-0.040	&	-0.100\\
			Densities&No&
			0.001	&	-0.004	&	-0.046	&	0.007	&	0.004	&	-0.132\\
			\hline
			Variational& Yes&
			0.006	&	0.011	&	0.008	&	0.017	&	0.028	&	0.020\\
			Coulomb&No&
			0.005	&	0.010	&	0.007	&	0.015	&	0.026	&	0.018\\
			\hline
			Variational& Yes&
			0.005	&	0.012	&	0.010\\
			Polarization&No&		
			0.005	&	0.011	&	0.006\\
		\end{tabular}
	\end{ruledtabular}
	\label{Table3}
\end{table*}

The ParaGauss and ADFT methods are both referred to as ``Variational Coulomb" in Tables~\ref{Table1}-\ref{Table3}. Because the only difference between these two approaches is that ParaGauss is constrained, ``Yes" under the ``Constraint" column refers to ParaGauss and ``No" refers to ADFT.  Results obtained with our new method which variationally fits the total Coulomb and XC energy are labeled ``Variational Densities".  The method referred to as ``Numeric" uses Coulomb fitting to compute the Coulomb energy and the orbitals to numerically compute the XC energy.  All energies in Tables~\ref{Table1} and \ref{Table2} are relative to the unconstrained numeric approach.

Because of spin polarization in Mn, there are roughly twice as many variational parameters as in Zn.   In the method labeled ``Variational Polarization" in Table~\ref{Table2}, the two independent generalized densities that were fitted are the spin polarization and total density.   Because the spin-polarization only affects XC, it was fitted using a negative-definite metric matrix.  

The fits of both the model potential of ParaGauss and ADFT make the Coulomb energy and not the total EE stationary.  The first-order energy required to make the EE robust is not computed in those second-generation variational-fitting codes, but it is fairly easy to add, and we have done so in this work.  For non-spin-polarized systems, the methods are almost identical and quite accurate after the first-order correction is added.  Table~\ref{Table3} gives the robust correction to the fitted XC energies.

Integration was performed with the parameter-free 80-point radial grid of K\"oster, et al. \cite{Koester2004-2} which was chosen because no energy changed from that obtained using a 70-point grid to the accuracy of the following tables.   Potentials and second derivatives of the XC energy were obtained with the libxc library \cite{Marques2012}.  The  VWN LDA\cite{Vosko1980} and the PBE \cite{Perdew1996,Perdew1996Err} and  BLYP \cite{Becke1988,Lee1988,Miehlich1989} GGA functionals were chosen.

Variational fitting was also implemented in a completely numerical fashion over the radial grid of 80 points, involving the inverse of an 80 by 80 matrix for each functional and atom.  This allowed for complete variational freedom in the fitted density at each point.   We found agreement between the fitted and exact densities to machine precision when this was done.  Thus variational density fitting works, but it is important to begin to understand the effects of incomplete basis sets.

The variational fitting equations were solved by taking a single Newton-Raphson step at each SCF cycle, which requires the inversion of a matrix with dimension equal to the number of fitting functions.  Because Coulomb-fitting typically includes a perturbative correction to the Coulomb matrix,\cite{Dunlap1982} it also requires inverting a matrix of the same size at each cycle.  Therefore, the computational costs of the two methods are roughly equivalent.

Because the Coulomb and XC energies have opposite signs, the Hessian of the EE is not positive definite, and the solution to the fitting equations is in general a stationary point, but not a minimum in the 17 dimensional space spanned by the Zn fitting functions, FIG.~\ref{Figure1}\,(c). Because the first derivative of the EE with respect to fitting coefficients is zero, the second derivative, and therefore, the magnitude of Hessian eigenvalues, is a rough measure of how much the EE changes in a given direction.

For both Mn and Zn, the largest Hessian eigenvalue is several orders of magnitude smaller than the absolute value of the smallest eigenvalue.  This can likely be explained by the fact that the Coulomb energy dominates the EE.

Figure~\ref{Figure1} contains two dimensional slices of the EE of Zn versus fitting coefficients as computed with VWN.  In FIG.~\ref{Figure1}\,(a) and (b), Coulomb fitting is initially employed, which makes the Coulomb energy and not the total EE stationary.  We can demonstrate this by noting that if Coulomb fitting made the total the EE stationary, the plots should appear to be centered at a stationary point, regardless of the particular two-dimensional subspace that we choose.

As with the total EE, we should expect that when the robust Coulomb energy is stationary, its  second derivative is a good measure of how much it changes along a given direction.  When varying the fitting coefficients along the two eigenvectors of the Coulomb matrix with the largest eigenvalues, Fig.~\ref{Figure1}\,(a), the biggest change in EE comes from the Coulomb energy.  In these directions, Coulomb fitting nearly maximizes the EE.

The effects of XC become more apparent along the Coulomb matrix eigenvectors with the smallest eigenvalues.  In these directions, the change in Coulomb energy is much smaller and the otherwise negligible contribution of XC plays a proportionally larger role.  In FIG.~\ref{Figure1}\,(b), it can be seen that the negative XC energy creates an EE minimum in these directions.  This minimum is also far from the center of the plot, which represents the original Coulomb-fitting solution.  Therefore, the Coulomb fit has not found a stationary point of the robust EE.

In our method, the fit is a true stationary point of the EE.  The fact that our fit finds a saddle point is illustrated in FIG.~\ref{Figure1}\,(c), which is plotted along two eigenvectors of the EE Hessian whose eigenvalues have opposite signs.

The saddle points we find are different from those described in the minmax procedure of K\"oster et al. \cite{Koester2009} because in that work, there is a separate Coulomb energy maximization with respect to $\bar{\rho}$ and total energy minimization with respect to the orbitals.  That problem separates into two convex optimizations in different sets of variables.  Our fitting equations are solved at a saddle point with respect to the fitting coefficients alone.  This differs from traditional quantum chemistry methods that do not use fitting, where pure minimization techniques can be employed.

Atomic Zn calculations used the Turbomole triple-zeta plus polarization (tzp) orbital basis \cite{Schafer1994}, contracted from a primitive 17s/10p/6d orbital basis.  The fitting basis set was created from the {\it s}-orbital basis set with every exponent doubled \cite{Dunlap1979,Dunlap1979-2}.  This basis is not optimized for Coulomb fitting and would not be expected to favor Coulomb over EE fitting.  Atomic Mn calculations used the refined \cite{Calaminici2007}  double-zeta polarization (dzp) deMon2k orbital basis  (contracted from 15s/9p/5d).

With a fitted Coulomb energy and a numerical treatment of the exact, orbital-derived XC energy, the total PBE energy for Zn is -1779.12123\,Hartree.  Even tempering the fitting basis gives an energy that is 0.0023 H below that using the orbital-derived fitting basis set.  Thus it is less accurate, because the exact Coulomb energy bounds the fitted Coulomb energy from above.   An identical calculation can be performed using the smaller, 15s/9p/5d primitive, DGauss DZVP2 orbital basis set \cite{Andzelm1992}.  Its energy lies 0.3315\,H higher.  These energy differences set an appropriate energy scale for this work.   Any method with an error of less than the middle ground of $\pm$0.03\,H would be expected to provide useful transition-metal quantum chemistry and is used as an arbitrary standard of accuracy.

All errors in the total energy are below our accuracy standard.  The closest any calculation comes to crossing that threshold is the constrained, BLYP, variational densities fit for Zn, which is has an error of -0.016\,H.  In all other cases, the error is much smaller and our method performs nearly identically to Coulomb fitting.  The error is an order of magnitude below our standard and approximately the same as the difference between the even tempered and orbital-derived basis sets.

For BLYP, the first order XC corrections with our method, the first two rows of  Table~\ref{Table3}, are greater in magnitude than the accuracy standard.  The first order correction in Zn with BLYP is particularly large.  All other values in Table~\ref{Table3} are much smaller but should not be ignored.

We have, for the first time, defined a universal robust expression for a fitted EE, Eq.~(\ref{Eq2}), and determined the corresponding coupled set of fitting equations, Eq.~(\ref{Eq3}), for all electronic-structure methods.  In DFT, it simplifies previous results for variational fitting of the Coulomb and Slater XC potentials, which involve separate fits of the density and Slater exchange potential \cite{Dunlap2011} at the price of numerical integration.  The largest errors in this preliminary study of full variational fitting in DFT are those needed to make the fitted energy robust.  In all cases, the first-order XC error is significant.

Of the methods that use fitting to compute XC, only ours makes the entire energy stationary with respect to $\bar{\rho}$.  The fact that the fitting equations are often satisfied at an energy saddle point means that the properties of our new variational principle are somewhat different than those of traditional quantum chemistry methods.  As good basis sets and zeroth-order fits of the non-negative density exist, it is possible that Newton-Raphson will always work, as it does with Slater exchange.  Our preliminary investigation has demonstrated that this method can be employed effectively for DFT.

\begin{acknowledgments}
	We thank Marcin Dulak for help in linking libxc on a Mac.  This work is supported by the Office of Naval Research, directly and through the Naval Research Laboratory.  M.C.P. gratefully acknowledges an NRC/NRL Postdoctoral Research Associateship.
\end{acknowledgments}

\bibliography{citations}
\end{document}